# Behavior of ZnO-coated alumina dielectric barrier discharge in atmospheric pressure air

Meng Li, Hai Dong, Xiaoping Tao

Department of physics, University of Science and Technology of China, Hefei, Anhui, China

**Abstract**: *A complete investigation of the discharge behavior of dielectric barrier discharge device using ZnO-coated dielectric layer in atmospheric pressure is made. Highly conductive ZnO film was deposited on the dielectric surface. Discharge characteristic of the dielectric barrier discharge are examined in different aspects. Experimental result shows that discharge uniformity is improved definitely in the case of ZnO-coated dielectric barrier discharge. And relevant theoretical models and explanation are presented to describing its discharge physics.*

**Key words**: ZnO-coated surface, Dielectric barrier discharge, Uniform discharge, Lissajous figure, Atmospheric pressure air, Scanning electron microscope, Discharge power

## 1. Introduction

The generation of uniform discharges in open air is one of the most remarkable researches in the plasma field. Despite most researches about discharges with additional flowing gases and vacuum chamber, uniform and stable discharges in the open air have many advantages, which can be used in many plasma industries, such as surface modification, deposition, cleaning, etc, and there have been some reports about uniform glow or glowlike discharge using a dielectric barrier discharge (DBD) structure in atmospheric pressure [1-5].However, in the case of open air discharge, oxygen molecules consisting of 21% in air quench nitrogen metastable species, so atmospheric discharges exist as streamer, nonuniform microdischarges[9]. So it is important to minimize this quenching effect and make the discharges in the open air uniform and stable.

This letter reports how ZnO-coated dielectric barrier improves plasma uniformity and influence the discharge behavior of DBD profoundly from different aspects. With ZnO thin film on alumina using rf magnetron spatter causes about a million higher surface conductivity than bare alumina surface [9], our experimental results will show what effects the change of surface conductivity has on discharge behavior including discharge uniformity and discharge power in open air.

## 2. Experimental set up

(1) Preparation of dielectric barrier

Before the experiment, pure alumina powder was compressed into thin plates under pressure 10 times of the atmospheric pressure and then annealed for 8 hours. The Alumina plates have a thickness of 2.32 mm and a diameter of 13 mm.

After that half of the Alumina plates were deposited with ZnO thin film on their surface, using rf magnetron sputtering with a sintered ZnO target (99.99%).Fig 1 shows SEM images of two kinds of barriers' surface structure at the same scale (×200000) before discharge. The surface of the two kinds of materials is both rough, but there are densely distributed ZnO spots on the surface of Alumnia after deposition, forming a 20-nm thin ZnO film without bringing too much change to the substrate roughness (shown in Fig (b)).

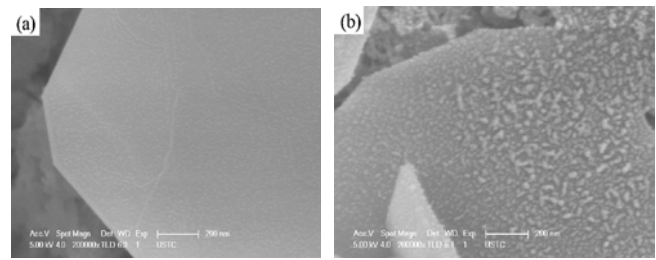

*Fig1: The SEM image of the surface structure of Alumina barrier (a) and the ZnO-coated alumina barrier (b)*

(2)The discharge apparatus

The DBD reactor has a parallel-plate type electrode configuration. The two electrodes are all highly-conductive cylindrical graphite electrodes, and the barrier is adhered to the graphite electrodes to form a DBD device. The circuit of DBD device is shown in Fig 2:

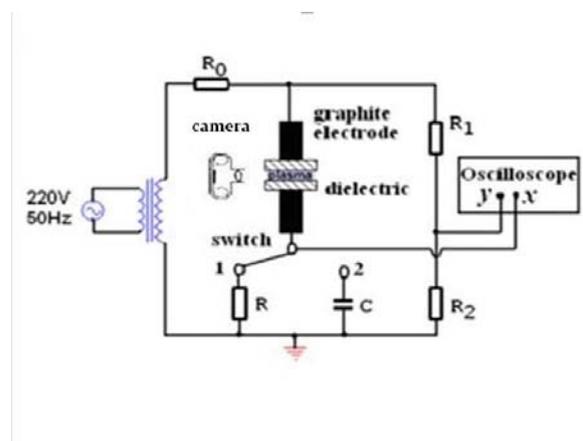

*Fig 2: Schematic diagram of the experimental set-up，and the parameter of the circuit is:*

$R_0 = 16K\Omega$, $R = 100\Omega$, $R_2 = 0.1M\Omega$, $R_1 = 100M\Omega$,

$C = 0.1\mu F$

The circuit can measure discharge current as well as discharge power:

When the switch connect to 1, the figure of discharge current and applied voltage on DBD device can be measured in the oscilloscope. When the switch connected to node 2, the Lissajous figure can be attain from the X-Y Mode on the oscilloscope. Since Voltage Vc on the capacitance can be measured. So

$$I = C\frac{dV_c}{dt}$$

Then the discharge power is

$$P = \frac{1}{T}\int_0^t VIdt = \frac{C}{T}\int_0^t V\frac{dV_c}{dt}dt = fC\oint VdV_c \quad (1)$$

And $C\oint VdV_c$ is the area of Lissajous Figure, thus the discharge power can be calculated [11].

### 3. Experiment results and discussion

During the experiment, The DBD device is exposed in the open air. The 220V, 50Hz sinusoidal voltage is connected in the circuit after a transformer, so that the voltage on the DBD can vary from 0 to 9.5KV. As voltage was applied, micro discharge was generated in the zone between two planar electrodes. For the two kinds of dielectric barrier, we keep the pressure $10^5$ Pa and the air gap width 1mm. Then their discharge current and power are measured and the figures are recorded on the oscilloscope.

#### 3.1 Improvement of discharge uniformity

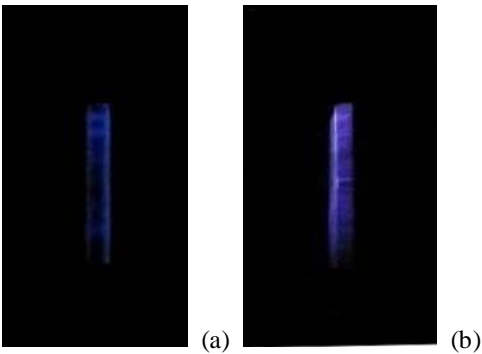

(a)  (b)

*Fig 3: The photo of Alumnia(a) and ZnO-coated alumina(b) DBD discharge under the same voltage in a darkroom(taken by 300,000 pixels digital camera)*

Fig 3(a) and Fig 3(b) are the discharge photos under the same applied voltage (7000V). Fig 3(b) shows that the discharge at the ZnO-coated surface seemed uniform, stable and a larger, glow like discharge area is formed in the gap; while the discharge of bare Alumina have many separate discharge channels in the gap just as filamentous discharge.

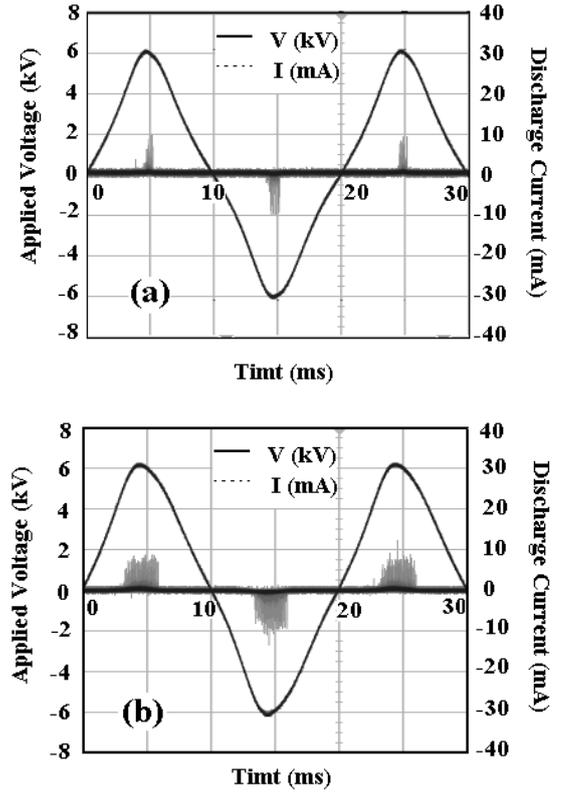

*Figure 4: The discharge current and applied voltage wave forms of Alumina (a) and Alumina coated with ZnO (b) under the same voltage (6000V)*

Then wave forms of discharge current and applied voltage were recorded on the oscilloscope. Fig 4 is a comparison of the current wave forms between Alumina barrier and ZnO-coated Alumina barrier under 6000V. Fig 4(b) indicates that more micro discharges were produced on ZnO-coated Alumina under the same circumstance. Though the intensity of one single micro discharge is about the same (4mA), the amount of single micro discharge on ZnO-coated surface is about 5 times of the Alumina.

It is clear that the surface structure is not largely affected after the deposition of the ZnO film as shown in Fig 1, so other possibilities to change surface properties such as capacitance, secondary electron emission coefficient, and surface conductivity may be responsible for the change of discharge behavior. The total capacitance of Alumina and ZnO film is

$$\frac{1}{C_{total}} = \frac{1}{C_{Al_2O_3}} + \frac{1}{C_{ZnO}} = \frac{1}{S}\cdot(\frac{d_{Al_2O_3}}{\varepsilon_{Al_2O_3}} + \frac{d_{ZnO}}{\varepsilon_{ZnO}}) \approx \frac{1}{C_{Al_2O_3}} \quad (2)$$

Since the ZnO film is just about 20nm, the capacitance is not largely affected. For the atmospheric pressure discharge, the mean free path of positive ions is too short to accelerate enough to generate secondary electron emission [7].

However, it has been demonstrated by Novak and Bartnikas [12-14] that the presence of surface charge on an electrode surface modifies radial distribution of the electron density in the gap volume and, consequently, affects the discharge mechanism itself. The effect of charge spreading along the dielectric surface may be ascertained by assuming ohmic behavior to prevail and deploying the analytical solution proposed by Somerville and Vidaud [6]. According to their model, the radial field due to a surface point charge is defined by

$$E_r(r,0,t) = \frac{Q_p r}{4\pi\varepsilon_0 \varepsilon_r (r^2 + v^2 t^2)^{3/2}}\bigg|_{t>0} \quad (3)$$

Where is the constant velocity defined by

$$v = \frac{1}{2R(\varepsilon_0 \varepsilon_r)} \quad (4)$$

where R is the surface resistivity. In this reference, when using a material with high surface conductivity and low R, the radial field between Qp and r with time is decreased fast, and this result means fast charge expansion by surface conduction.

At discharge onset when applied voltage researches breakdown voltage, the charges will be deposited on the dielectric barrier surface by the ensuing discharge. If the surface conductivity has a low value, a given discharge event will not affect the charge distribution over large portions of the multitude of discharge sites, thus the voltage will drop to a remnant value V at that site and the discharge will cease on account of the reduced field. This results in localized space charge formation in the gas, just like the filamentous type discharges with many separate micro discharges, which has been revealed with numerical simulation[10].

Nevertheless, in the case of ZnO-coated barrier, the surface conductivity is 1,000,000 times higher than the alumina only electrode; the lifetime of micro discharges in DBD is generally tens of nanoseconds so charges can spread within a discharge life[9].Since electrical charges from the discharge site can now move more easily over the surface of the dielectric barrier, when the charges are deposited on the dielectric barrier surface through the discharge channel, the accompanying field reduction will affect a larger area. As a result, the remnant voltage V, at the site after discharge will still be comparable to the breakdown voltage Vb and will not cause extinction of individual discharges; instead, the discharges persist at a lower overvoltage level as other discharge sites ignite, resulting in long-lasting channel and more microdischarges appears from different sites (shown in figure 4), eventually the discharge looks like a faint glow filling the entire gap [7] (as the figure 3 shows).In all, The charge spreading over a wide area due to high surface conductivity enhances the discharge uniformity and cause the disappearance of the intense pulse discharges, and leads to pseudoglow discharges [12-14].

### 3.2 Improvement of discharge power

The discharge power is an important parameter in discharge since it represents the efficiency of gas discharge, and usually the different power indicates the different state of plasma. From equation (1) we know that the discharge power can be measured though measurement of the area of Lissajous figures. One group of the Lissjous figures we got is shown Fig 5.

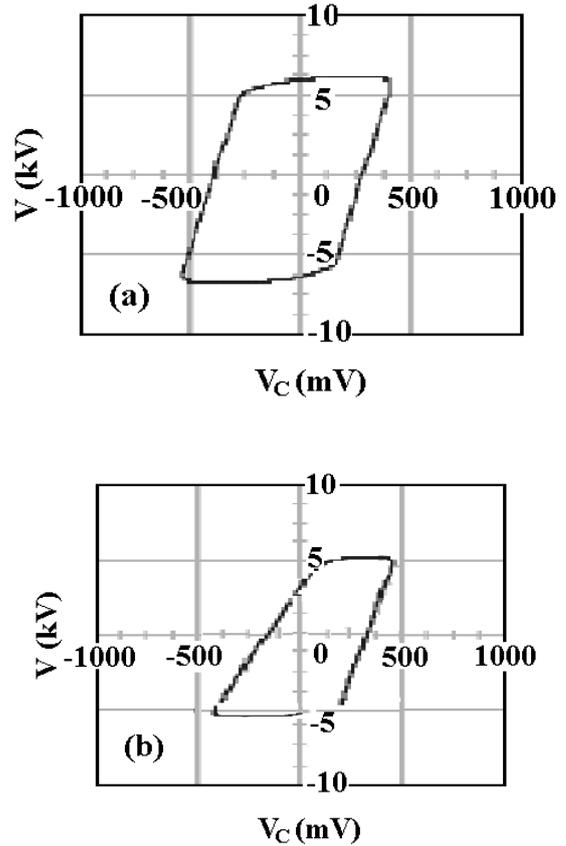

*Figure 5: The discharge Lissjous figure of Alumina coated with ZnO(a) and Alumina (b) under the same voltage (6000V)*

Just like typical Lissjous Figures of DBD, the two figures are parallelogram. The larger area of Lissjous Figure (Fig 5(a)) represents ZnO-coated Alumina has a larger discharge power under the same voltage (6000V).

We kept the air gap and pressure the same and changed the applied voltage on DBD device from 5000V to 9500 V (their Lissjous figure was recorded every 500V), after calculation (according to Equation 1) we got their discharge power under different voltage (shown in table1 and Fig 6):

| Applied voltage U/kV | Discharge power of Alumina barrier/W | Discharge power of ZnO-coated Alumina /W |
|---|---|---|
| 9.5 | 0.425 | 1.85 |
| 9 | 0.3125 | 1.45 |
| 8.5 | 0.275 | 1.2 |

| 8 | 0.2375 | 0.7 |
| 7.5 | 0.2 | 0.55 |
| 7 | 0.15 | 0.4 |
| 6.5 | 0.0625 | 0.1 |
| 6 | 0.025 | 0.05 |
| 5.5 | 0.00625 | 0.0075 |
| 5 | 0.0004 | 0.0005 |

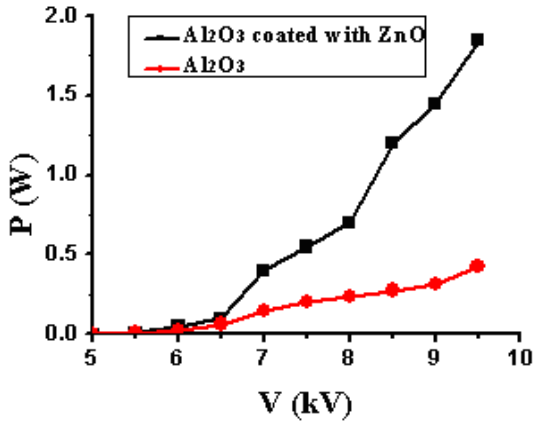

*Figure 6: the discharge power of Alumina and ZnO-coated Alumina under different voltages*

Fig 6 shows that both ZnO-coated Alumina and bare Alumina begin to discharge under the voltage of about 5500V, and then the power grows rapidly. However, the discharge power of ZnO-coated barrier is much higher than Alumina (about 4 times higher under 9000V). It is because ZnO-coated barrier have more microdischarges in one discharge period (as the results in 3.1 shows). Though the current of a single discharge pulse is about the same, more charges transport through the gap via the channel and eventually the discharge power is improved significantly.

However, too high surface conductivity, such as the case of metal-to-metal atmospheric pressure discharge will not enhance the discharge uniformity and glow discharge because of insufficient current limiting with too high conductivity.

## 4. Summary

In summary, ZnO-coated Alumina and Alumina are employed as barrier dielectrics. DBD devices are exposed in atmospheric open air, and SEM image, the discharge voltage and current waveforms and Lissajous figure were measured in this experiment. The experiment result indicates that uniform plasma discharge in air could be generated by controlling electrical properties of the DBD barrier surface. A highly conductive ZnO film deposited on the alumina surface promotes the charge spreading by radical electric field, and is responsible for more microdischarges produced during discharge.